\documentclass[aps,twocolumn,superscriptaddress,longbibliography,prl,10pt]{revtex4-1}%
\usepackage{amsfonts}
\usepackage{amsmath}
\usepackage{amssymb}
\usepackage{graphicx}
\usepackage{dcolumn}
\usepackage{color}
\usepackage{natbib}
\usepackage{hyperref}
\usepackage{bm}

\begin{document}

\flushbottom

\title{Subsurface bending  and reorientation  of tilted vortex lattices in the bulk due to Coulomb-like repulsion at the surface} 

\author{E. Herrera}
\affiliation{Laboratorio de Bajas Temperaturas y Altos Campos Magn\'eticos, Unidad Asociada UAM/CSIC, Departamento de F\'isica de la Materia Condensada, Instituto de Ciencia de Materiales Nico\'las Cabrera, Instituto de F\'isica de la Materia Condensada, Universidad Aut\'onoma de Madrid, E-28049 Madrid, Spain}

\author{I. Guillam\'on}
\affiliation{Laboratorio de Bajas Temperaturas y Altos Campos Magn\'eticos, Unidad Asociada UAM/CSIC, Departamento de F\'isica de la Materia Condensada, Instituto de Ciencia de Materiales Nico\'las Cabrera, Instituto de F\'isica de la Materia Condensada, Universidad Aut\'onoma de Madrid, E-28049 Madrid, Spain}

\author{J.A. Galvis}
\affiliation{Laboratorio de Bajas Temperaturas y Altos Campos Magn\'eticos, Unidad Asociada UAM/CSIC, Departamento de F\'isica de la Materia Condensada, Instituto de Ciencia de Materiales Nico\'las Cabrera, Instituto de F\'isica de la Materia Condensada, Universidad Aut\'onoma de Madrid, E-28049 Madrid, Spain}
\affiliation{Departamento de ciencias naturales, Facultad de ingenieria, Universidad Central, Bogot\'a, Colombia.}
\affiliation{National High Magnetic Field Laboratory, Florida State University, Tallahassee, FL 32310, USA.}

\author{A. Correa}
\affiliation{Laboratorio de Bajas Temperaturas y Altos Campos Magn\'eticos, Unidad Asociada UAM/CSIC, Departamento de F\'isica de la Materia Condensada, Instituto de Ciencia de Materiales Nico\'las Cabrera, Instituto de F\'isica de la Materia Condensada, Universidad Aut\'onoma de Madrid, E-28049 Madrid, Spain}
\affiliation{Instituto de Ciencia de Materiales de Madrid, Consejo Superior de Investigaciones Cient\'ificas, CSIC, E-28049 Madrid, Spain}

\author{A. Fente}
\affiliation{Laboratorio de Bajas Temperaturas y Altos Campos Magn\'eticos, Unidad Asociada UAM/CSIC, Departamento de F\'isica de la Materia Condensada, Instituto de Ciencia de Materiales Nico\'las Cabrera, Instituto de F\'isica de la Materia Condensada, Universidad Aut\'onoma de Madrid, E-28049 Madrid, Spain}

\author{S. Vieira}
\affiliation{Laboratorio de Bajas Temperaturas y Altos Campos Magn\'eticos, Unidad Asociada UAM/CSIC, Departamento de F\'isica de la Materia Condensada, Instituto de Ciencia de Materiales Nico\'las Cabrera, Instituto de F\'isica de la Materia Condensada, Universidad Aut\'onoma de Madrid, E-28049 Madrid, Spain}

\author{H. Suderow}
\affiliation{Laboratorio de Bajas Temperaturas y Altos Campos Magn\'eticos, Unidad Asociada UAM/CSIC, Departamento de F\'isica de la Materia Condensada, Instituto de Ciencia de Materiales Nico\'las Cabrera, Instituto de F\'isica de la Materia Condensada, Universidad Aut\'onoma de Madrid, E-28049 Madrid, Spain}

\author{A. Yu. Martynovich}
\affiliation{Ames Laboratory and Department of Physics \& Astronomy, 
Iowa State University, Ames, Iowa 50011, USA}

\author{V. G. Kogan}
\affiliation{Ames Laboratory and Department of Physics \& Astronomy, 
Iowa State University, Ames, Iowa 50011, USA}

\date{\today}

\begin{abstract}
We  study vortex lattices (VLs) in  superconducting  weak-pinning platelet-like crystals of $\beta$-Bi$_{2}$Pd in  tilted fields with a Scanning Tunneling Microscope. We show that vortices exit the sample perpendicular to the surface and are thus bent beneath the surface. The structure and orientation of tilted VL in the bulk are, for large tilt angles, strongly affected by Coulomb-type intervortex repulsion at the surface due to stray fields.
\end{abstract}

\maketitle

Vortices in  superconductors are predicted to bend in order to orient  perpendicular to  boundaries between  superconducting and normal areas or to the interface with vacuum   \cite{Brandt93,Martynovich93}. Physically, the bending is due to the vortex supercurrent loops which, on one hand, should be parallel to the surface, while on the other,  tend to be in plane perpendicular to the vortex axis (in isotropic materials). To the best of our knowledge, such bending has never been verified experimentally.  We provide Scanning Tunneling Microscopy (STM) data that cannot be interpreted in any other way---vortices exit the sample perpendicular to the  surface. 
Our data also show that in weak-pinning crystals the   VL structure and orientation are affected not only by the intervortex interactions in the bulk, but also by interactions  at the sample surface.

We study vortex lattices (VLs) at the plane surface of platelet-like   single crystals of $\beta$-Bi$_2$Pd with  $T_c=5\,$K \cite{Imai12,Alekseevski54}. The crystal  is tetragonal, although the Fermi surface has sheets of mixed orbital character that lead to a near-isotropic macroscopic behavior \cite{Shein13,Sakano15,Coldea16}. Upper critical fields along the basal plane and perpendicular to it differ by barely 25\% from $H_{c2,ab}=$ 0.7\,T to  $H_{c2,c}=$ 0.53\,T (at low temperatures), leading to    coherence lengths $\xi_{c}=19$\,nm and $\xi_{ab}=24$\,nm \cite{Kacmarcik16,Herrera15}. Estimates of the penetration depths from the data on the lower critical field give $\lambda_{ab}=105$\,nm and $\lambda_{c}=132$\,nm \cite{Kacmarcik16}.

We use a home-built STM/S attached to the dilution refrigerator inserted in a  three axis vector magnet reaching 5\,T in the $z$ direction and 1.2\,T  for  $x$ and $y$  \cite{Suderow11,Galvis15}.  $\beta$-Bi$_2$Pd crystals   (3 $\times$ 3 $\times$ 0.5 $mm^3$) are mounted with the $c$-axis   along the $z$ direction of the magnet. The other two crystalline orientations with respect to $x$ and $y$ of the magnet are found by scanning the surface with atomic resolution to find the square Bi lattice as outlined in \cite{Herrera15},  where the crystal   growth is also described. The surface consists of large atomically flat areas of several hundreds nm in size, separated by step edges  \cite{Herrera15}. We use an Au tip cleaned and atomically sharpened in-situ by repeated indentations onto  Au sample \cite{Rodrigo04}.  VL images are obtained by  mapping  the zero-bias conductance normalized to voltages above the superconducting gap  \cite{Guillamon08}. All measurements are done at $T=150$ mK. Data are usually taken within field-cooled protocol, although, due to  weak vortex pinning of this material, we find the same results when changing the magnetic field at low temperatures. No filtering or image treatment is applied to the conductance maps shown below.

 VLs in  fields along $c$ are hexagonal up to  $H_{c2}$    with one of the VL vectors along $a$ or $b$   of the tetragonal crystal \cite{Herrera15}. This gives a two-fold degeneracy in the VL orientation \cite{Granz16}. Hence we observe domains of differently oriented VLs; the spatial distribution of domains is random and determined by the pinning landscape. In some tetragonal materials in fields along $c$,  the four-fold-symmetric nonlocal corrections to the London theory modify the isotropic repulsion and lead to two degenerate rhombic VLs which at large fields transform to the  square VL  \cite{Eskildsen98,Yethiraj99}. One of the requirements for the nonlocal corrections to work is a large Ginzburg-Landau parameter $\kappa$ \cite{Kogan97}. In our crystals, $\kappa\sim 4$ and we do not observe VL transitions.

	\begin{figure}[h]
	\includegraphics[width=0.48\textwidth]{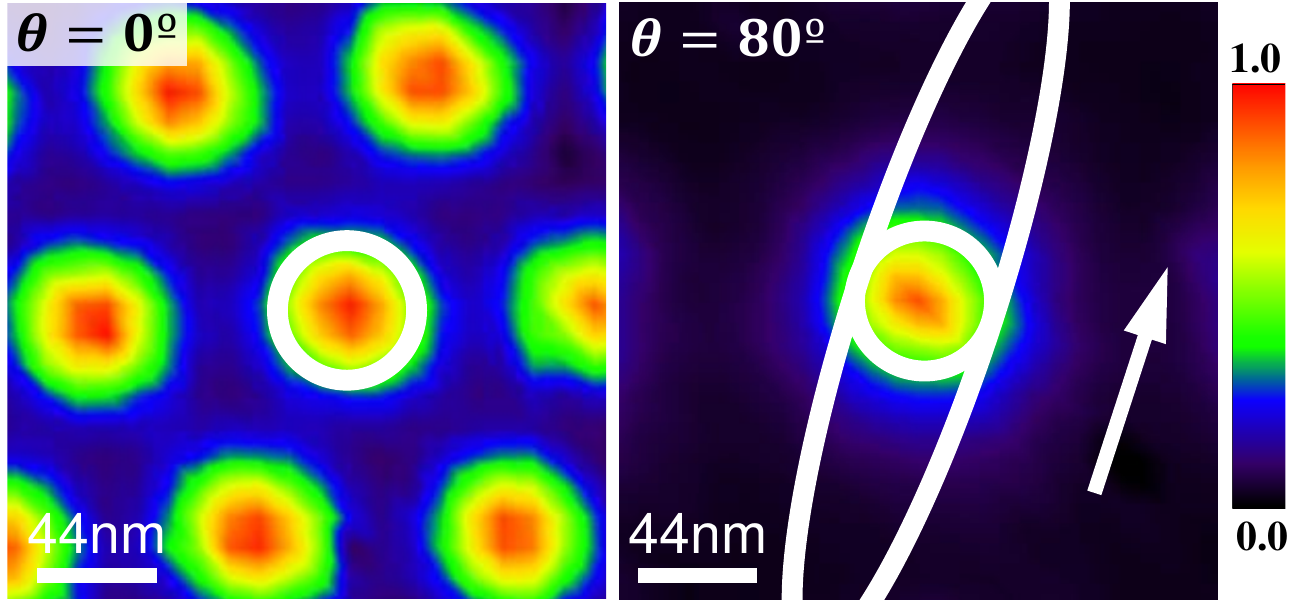}
	\caption{Zero-bias conductance maps for $H=0.3\,$T, $T=0.15\,$K; $\theta = 0^{\circ}$ (left) and $\theta=80^{\circ}$ (right).
	Vortex cores in both cases are roughly circles  of the same size. White circles have radii  $  \sim 24$ nm. The white ellipse has the major semi-axis of $24/\cos\theta$\,nm $= 138$\,nm. The  arrow indicates the tilt direction. Color bar represents the normalized tunneling conductance for both images. }
	\label{fig1}
	\end{figure}
 
Let us consider  the   vortex core shapes at the surface.
At $\bm H\parallel \bm c$, the cores have a circular shape of a size   
  $ \xi_{ab} \approx 24$\,nm at 0.3 T (see discussions \cite{Herrera15,Fente16}) shown in Fig.\,\ref{fig1} as a white circle. If the vortex in a tilted field would have arrived to the surface being straight  without any bending, the expected core shape at the surface would be an ellipse with the minor and major semiaxes of 24\,nm and $24/\cos\theta$ \,nm ($\theta$ is the angle between $\bm H$ and $\bm c$). For $\theta=80^\circ$ we would obtain the white ellipse shown in the right panel of Fig.\,\ref{fig1}. Instead we find  the vortex core of the same shape and size as for $\bm H$ normal to the surface as shown by the circle at the right panel of Fig.\,\ref{fig1}. Thus, our images   show that vortices must bend under the surface to exit the sample being perpendicular to the surface.

Vortex bending is expected to occur over a length of the order of the penetration depth $ \lambda \approx 100$\,nm \cite{Brandt93,Martynovich93}, which is   large relative to the core size of $ 24$\,nm. Hence, we do not expect that the electronic density of states at the surface is influenced by the bent part of the vortex deep underneath.

\begin{figure}[t]
\includegraphics[width=0.45\textwidth]{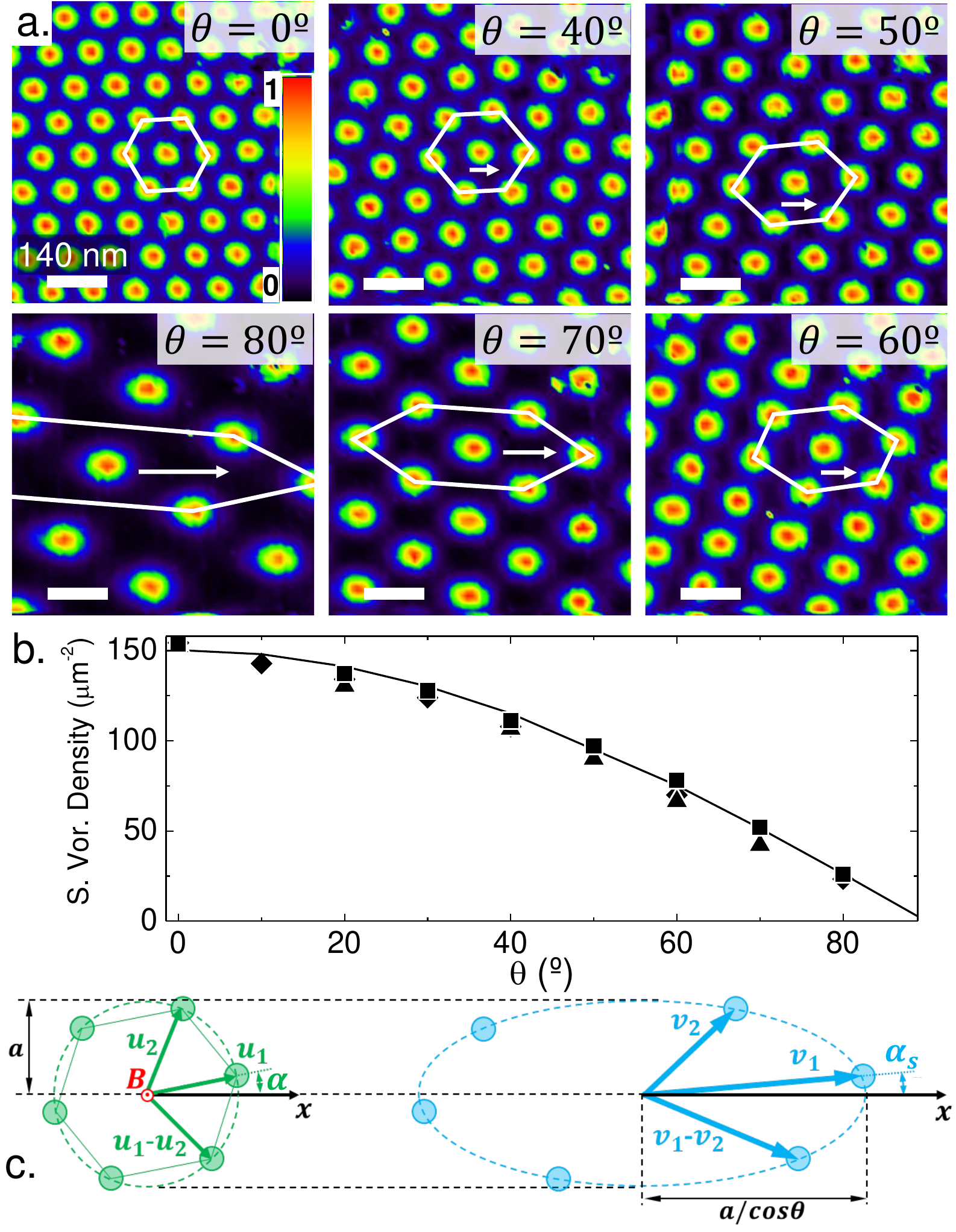}
\caption{(a) Zero-bias  density of states showing VLs for a few tilt angles $\theta$. The color scale  on the top-left panel  represents the normalized conductance.  White arrows indicate   tilt directions.    White distorted hexagons are  projections onto the surface of  hexagonal VLs in the bulk.   (b) Density of vortices  at the surface   vs $\theta$ and the normal component of the magnetic induction $ H\cos\theta$ (solid curve); different symbols represent different experiments. We use large fields, so that  the difference between the applied field $\bm H$ and the magnetic induction $\bm B$  can be disregarded. 
(c) Schematic representation of the bulk VL in the vortex frame (left panel) with lattice vectors $u_1$ and $u_2$ and its surface projection (right panel) with lattice vectors $v_1$ and $v_2$ for the tilt along $x$.
}
\label{fig2}
\end{figure}

The surface  VLs are shown in Fig.\,\ref{fig2}(a)   for a few   tilts   $\theta$. 
The panel 2(b) shows that the density of vortices at the surface goes  as $\cos\theta$,  as expected. 

As mentioned, the material of our interest   is nearly isotropic. The model we offer to describe  VLs in tilted fields  is {\it isotropic}. The model predictions, by and large,  agree  with the STM data.  Within this model, the VL in the vortex frame of an infinite sample is hexagonal and {\it degenerate}: the angle $\alpha$ shown on the left   of Fig.\,\ref{fig2}(c) can be taken as the degeneracy parameter. The circle where all   nearest neighbors   are situated has a radius $a$ fixed by the flux quantization, $  a^2 =  2\phi_0/\sqrt{3} H $. We use the vortex coordinate frame $(x,y,z)$ with  $z$  along the vortex direction and the $x$ axis in the tilt plane. For a given $\alpha$, the VL unit cell vectors (in units of $a$) are  
  \begin{eqnarray}
  {\bm u}_1= [\cos\alpha, \sin\alpha],\,\,\,
    {\bm u}_2 = \left[\cos\left(\alpha+\frac{\pi}{3}\right), \,\sin\left(\alpha+\frac{\pi}{3}\right)\right].\nonumber\\
\label{u}
\end{eqnarray}
In a sample much thicker than $\lambda$,  the VL structure in the  bulk is dominated by the bulk interactions,   i.e. the VL is still hexagonal. However, the degeneracy is removed in tilted fields by the surface contribution to the interaction. 
 
To evaluate this contribution, we  note that due to subsurface bending the point of vortex exit is shifted relative to the exit point of a straight vortex. In small fields when the vortices are well separated, each one will experience the same shift. We assume that shifts are the same also in fields of our interest. In particular, this implies that the density of bent vortices at the surface is the same as if vortices were straight; this is consistent with the   macroscopic boundary condition for the normal component of the magnetic induction $H\cos\theta$. Hence, the arrangement of vortices at the surface is just shifted relative to the VL which would have been there without subsurface bending. Then, considering the VL structure, one can disregard the bending and the bulk nearest neighbors will be situated at the  cros-section of a circular cylinder of radius $a$ with the flat surface, i.e. at the ellipse with semi-axes $a/\cos\theta$ and $a$, the right panel of  Fig.\,\ref{fig2}c. 
   Taking again the $x$ axis of the surface frame in the tilt plane, one obtains new unit cell vectors at the surface  (in units of $a$): 
     \begin{eqnarray}
  {\bm v}_1= \left[\frac{\cos\alpha}{\cos\theta}, \sin\alpha\right],\,\,
    {\bm v}_2 =\left[\frac{\cos ( \alpha+ \pi/3)}{\cos\theta} , \sin\left(\alpha+\frac{\pi}{3}\right)\right].\nonumber\\
\label{v}
\end{eqnarray}
In particular, the angle $\alpha_s$ between $\bm v_1$ and ${\bm{\hat x}}$ is related to the parameter $\alpha$ by 
$   \tan\alpha_s = \tan\alpha \cos\theta$. 
All vortex positions at the surface   ${\bm R}_{mn} =m {\bm v}_1+n {\bm v}_2 $ 
  ($m,n$ are integers) can be expressed in terms of $\theta$ and $\alpha$ (or $\alpha_s$). 

Interaction of vortices at the surface is due to  stray fields out of the sample, which can be approximated by   point ``monopoles" producing the magnetic flux $\phi_0$ in the free space within the solid angle $2\pi$. The interaction energy of the vortex at the origin with the rest is
      \begin{eqnarray}
\frac{\phi_0^2}{4\pi^2}\sum_{m,n}\,^\prime \frac{1}{R_{mn}}\,,
\label{int}
\end{eqnarray}
where $\sum^\prime$ is over all   $m,n$ except $m=n=0$. The surface vortex density   is $H\cos\theta/\phi_0$, so that surface interaction per cm$^2$ is
       \begin{eqnarray}
 F_s&=& \frac{\phi_0H\cos\theta}{4\pi^2}\sum_{m,n}\,^\prime \frac{1}{R_{mn}}=\frac{3^{1/4}\phi_0^{1/2}H^{3/2}}{4\sqrt{2}\pi^2}\,S(\alpha,\theta),\qquad \label{Fs}\\
 S &=&\,\cos^2\theta\sum_{m,n}\,^\prime\Big\{[m\cos\alpha+n\cos( \alpha+ \pi/3)]^2\nonumber\\
&+&\cos^2\theta [m\sin\alpha+n\sin( \alpha+ \pi/3)]^2 \Big\}^{-1/2} .
\label{S}
\end{eqnarray}
It is readily checked that $\partial S/\partial\alpha=0$ at $\alpha=-\pi/6$ and $\pi/3$. The corresponding structures in the vortex frame are  hexagons, called hereafter $A$ and $A'$; in $A$ two out of six nearest neighbors are at $y$ axis, in $A'$ they are at $x$.

The sum $S$ is divergent and as such depends on the summation domain. We, however, are interested only in the angular dependence of $S(\alpha)$, because of its role in removing the VL degeneracy in the bulk. The angular  dependence arises mostly due to vortices in the vicinity of the central one, because the number of far-away vortices grows with the distance and their contribution to the interaction is nearly isotropic. Our strategy for evaluation  of $S(\alpha)$ is based on  the fact that the Coulomb interaction out of the sample is isotropic and therefore we can do the summation within a circle $R_{mn} < L$, where $L$ is  large enough to  include a few ``nearest-neighbor shells" of vortices surrounding the one at the origin. To provide a smooth truncation we add a factor $e^{-R^2_{mn}/L^2}$  to the summand of Eq.\,(\ref{S}).

Results of these calculations are given in Fig.\,\ref{f3} for $\theta\approx 80^\circ, 70^\circ$ and $60^\circ$;   smaller tilt angles are discussed in the supplemental material. 
   \begin{figure}[h]
\begin{center}
 \includegraphics[width=8.cm] {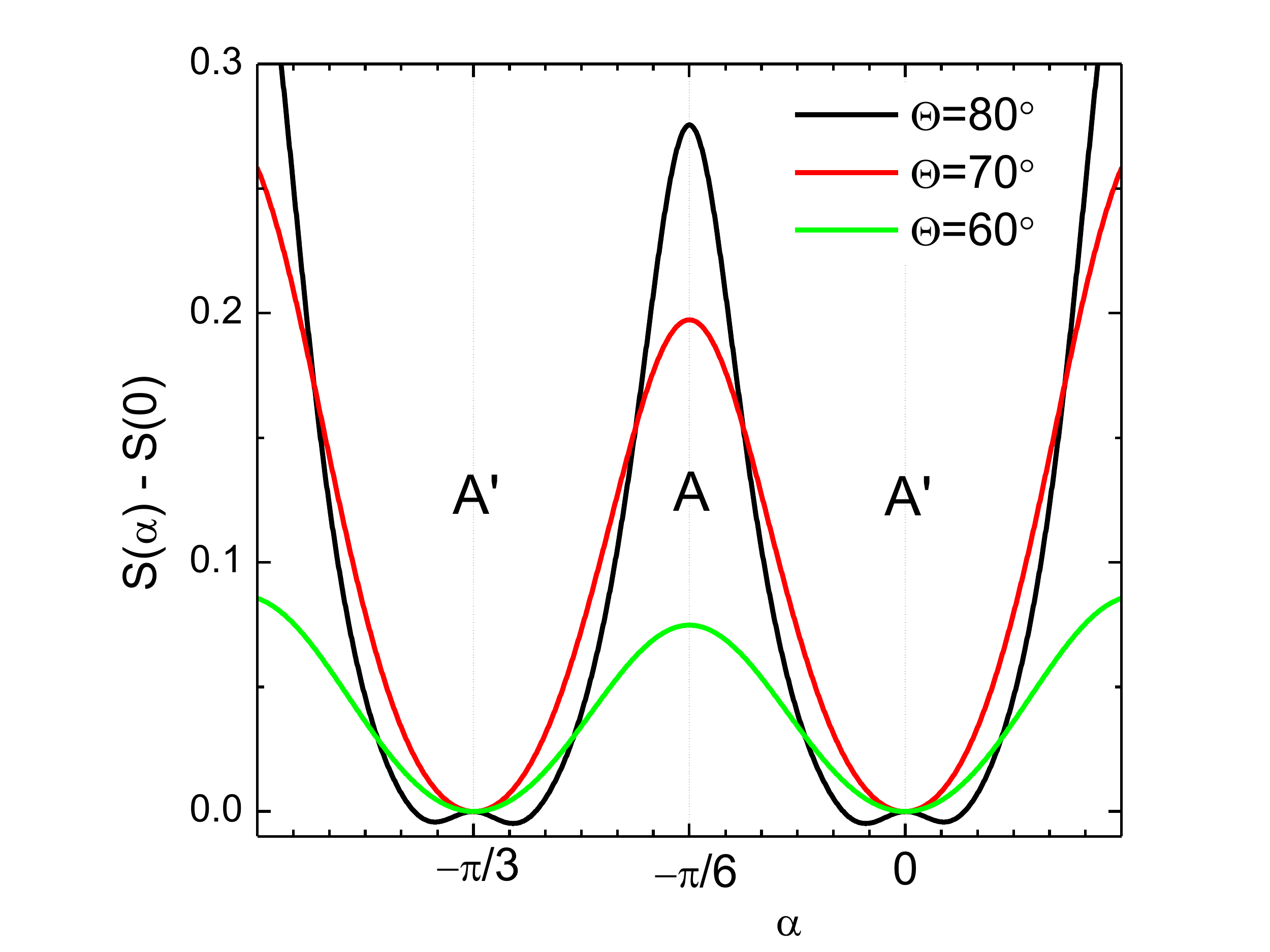}
\caption{ The surface energy angular dependence $S(\alpha)-S(0)$ for $\theta=  80^\circ (black), 70^\circ (red), 60^\circ (green)$    for $L= 4$  and  $-15<m,n<15$.  Curves $S(\alpha)$  are shifted vertically to have the same zero value  at $\alpha=0$.  The curve $\theta=80^\circ$ is in fact asymmetric relative to $\alpha=0$ with the absolute minimum at $\alpha\approx -0.1$.
}
\label{f3}
\end{center}
\end{figure}
Clearly, the structure $A$ ($\alpha=-\pi/6$) is unstable. The minimum energy for $\theta=  70^\circ$ and $60^\circ  $ is at $\alpha=0$, so that the preferred structure is  $A'$.  For $\theta=  80^\circ$ the minimum is shifted to $\alpha\approx -0.1$. These qualitative conclusions do not change if one takes a larger radius $L$ of the summation domain, notwithstanding the increase of the calculated surface energy $F_s$. 

To show that our model describes the data   well, and in particular 
to check again that the bulk hexagonal VL projects onto the surface as if the vortices were straight, one can follow the bulk nearest neighbors and their surface projections. 
The six nearest neighbors in the vortex frame correspond to the pairs $(m,n)$:  
      \begin{eqnarray}
(0,1), \,\,\,(1,0),\,\,\,(0,-1),\,\,\,(-1,0),\,\,\,(-1,1),\,\,\,(1,-1)\,.\qquad
\label{nn}
\end{eqnarray}
At the surface, these pairs mark six vortex positions situated at $\pm\bm v_1$, $\pm\bm v_2$, and $\pm (\bm v_1- \bm v_2)$, see the sketch in Fig.\,\ref{fig2}c.   These positions at the surface are not necessarily nearest, because at large tilt angles, they form a strongly stretched hexagon, whereas 
the position $(1,1)$, moves closer to the center.
 
Let us consider   $\theta=80^\circ$ for which according to Fig.\,\ref{f3} $\alpha\approx -0.1$ and evaluate    
distances $d_1=|\bm v_1|$, $d_2=|\bm v_2|$, and $d_3=|\bm v_1-\bm v_2|$. Skipping  algebra, we provide formulas for these distances in the supplemental material. In units of $a$ we obtain $(d_1,d_2,d_3)=(5.68,3.56,2.63)$. Direct measurements at the corresponding image at Fig.\,\ref{fig1} give $(d_1,d_2,d_3)\approx (6, 3.4, 2.4)$ in a good agreement with calculated values. Hence, the nearly degenerate hexagonal VL in the bulk bends as a whole when reaching the surface, just shifting the geometric projection of the tilted hexagonal bulk VL onto the surface.

\begin{figure}[h]
\begin{center}
 \includegraphics[width=8.cm] {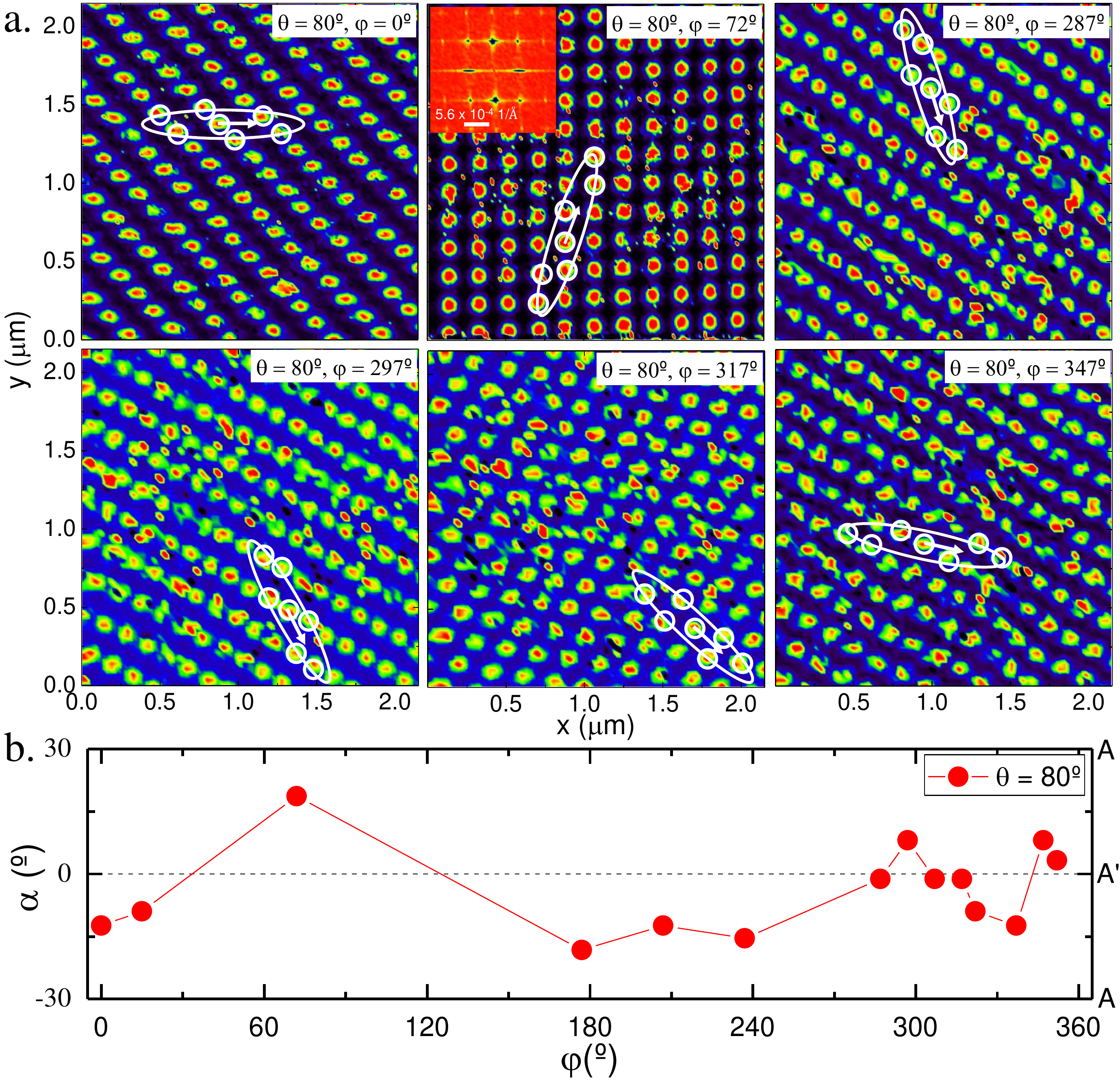}
\caption{ VLs for a fixed tilt $\theta=  80^\circ$ but different azimuthal angles $\varphi$ of the applied field. For $\varphi=72^\circ$ the lattice appears at the surface as a square (confirmed by the Fourier transform in the red inset), whereas for $\varphi=317^\circ$ it is nearly hexagonal. Note also that VLs of vortices strongly bent under the surface at high tilts $\theta$ are still very well ordered. 
 }
\label{f4}
\end{center}
\end{figure}

In Fig.\,\ref{f4}a we show results for a fixed polar angle (tilt) $\theta=80^\circ$ and several azimuthal angles $\varphi$ of the field $\bm B$. Interestingly, the surface vortex lattice shows different arrangements depending on $\varphi$. For $\varphi=72^\circ$,  the surface VL is a nearly perfect square, whereas for $\varphi=317^\circ$ it is nearly hexagonal. As we show in the supplementary material, the $\theta$ and $\varphi$ where we expect such a square vortex lattice within our model are close to what we observe experimentally.

In Fig. Fig.\,\ref{f4}b we show $\alpha$ for experiments changing the azimuthal angle $\varphi$ (for a polar angle of $\theta=80^\circ$). We observe that $\alpha$ varies around the orientation A'. The variations might be caused by the tetragonal symmetry, not included in our model or by weak pinning. 

Tilted VLs have been studied using STM and neutron scattering in anisotropic   2H-NbSe$_2$ \cite{Bending1999,Buzdin02,Koshelev05,Hess92,Hess94} and in materials with extreme anisotropies such as Bi$_2$Sr$_2$CaCu$_2$O$_{\delta}$.   In the latter, Josephson and pancake vortices form crossing lattices and VL properties are more involved than in our case  \cite{Bending1999,Buzdin02,Koshelev05}. The tilted VLs  in  2H-NbSe$_2$ are distorted hexagons 
\cite{Bolle93,Gammel94,Hess92,Hess94,Fridman11,Fridman13} whose  orientation, however,  disagrees  with theoretical expectations. It was proposed that vortex-induced  strain of the crystal might explain the data \cite{Kogan95}.  At  high tilt angles,  buckling transitions produce superlattices with chain-like vortex arrangements \cite{Hess92,Hess94}. Besides,  vortex core STM images have  star-like shapes  in fields along $c$ and acquire  comet-like tails in tilted fields with no evidence for vortices exiting the sample perpendicular to the surface \cite{Hess90, Galvis16}. Clearly, the results shown here are different, most probably due to the fact that $\beta-$Bi$_2$Pd is close to being isotropic, other than those systems.

We note also that high purity Nb crystals  were studied in tilted  fields \cite{Muhlbauer09}. 
The bulk VL shows a variety of transitions, including two-fold structures breaking the crystalline four-fold rotation symmetry and scalene unit cells. 
In pure Nb, however,   the Ginzburg-Landau $\kappa\sim 1$, the London approach does not apply, the microscopic theory becomes a necessity, and interpretation of VL structures is difficult.

In summary, we have studied   VLs in     tilted fields in $\beta$-Bi$_2$Pd, a  nearly isotropic superconductor. We  demonstrate that vortices exit the sample being  perpendicular to the surface, that necessitates the subsurface bending of vortex lines. We  find that intervortex Coulomb-like repulsion at the surface due to stray fields removes the degeneracy of the bulk hexagonal VLs thus fixing   the bulk VL orientation.  It is quite surprising to have a highly ordered VLs at the surface whereas under the surface all vortices are bent.
\\


Authors are grateful to P.C. Canfield for discussions and for having proposed growth of single crystals of $\beta-$Bi$_2$Pd and shown how to do that. E.H. is supported by the Departamento Administrativo de Ciencia, Tecnolog\'ia e Innovaci\'on, COLCIENCIAS (Colombia) Programa Doctorados en el Exterior Convocatoria 568-2012. I.G. is supported by the ERC (grant agreement 679080). This work was also supported by the Spanish MINECO (FIS2014-54498-R, MAT2014-52405-C2-02), by the Comunidad de Madrid through program NANOFRONTMAG-CM (S2013/MIT-2850) and by Axa Research Funds. We also acknowledge SEGAINVEX workshop of UAM, Banco Santander and COST MP1201 action, and the EU through grant agreements FP7-PEOPLE-2013-CIG 618321, 604391 and Nanopyme FP7-NMP-2012-SMALL-6 NMP3-SL 2012-310516.  V.K. is supported by the U.S. Department of Energy, Office of Science, Basic Energy Sciences, Materials Sciences and Engineering Division. The Ames Laboratory is
operated for the U.S. DOE by Iowa State University under Contract No. DE-AC02-07CH11358.  


\begin{thebibliography}{31}
\providecommand{\natexlab}[1]{#1}
\providecommand{\url}[1]{\texttt{#1}}
\expandafter\ifx\csname urlstyle\endcsname\relax
  \providecommand{\doi}[1]{doi: #1}\else
  \providecommand{\doi}{doi: \begingroup \urlstyle{rm}\Url}\fi

\bibitem[Brandt(1993)]{Brandt93}
E.H. Brandt.
\newblock Tilted and curved vortices in anisotropic superconducting films.
\newblock \emph{Phys. Rev. B}, 48:\penalty0 6699, 1993.

\bibitem[Martynovich(1994)]{Martynovich93}
A.~Martynovich.
\newblock Magnetic vortices in layered superconducting slabs.
\newblock \emph{Zh. Eksp. Teor. Fis.}, 105:\penalty0 912, 1994.

\bibitem[Imai et~al.(2012)Imai, Nabeshima, Yoshinaka, Miyatani, Kondo, Komiya,
  Tsukada, and Maeda]{Imai12}
Yoshinori Imai, Fuyuki Nabeshima, Taiki Yoshinaka, Kosuke Miyatani, Ryusuke
  Kondo, Seiki Komiya, Ichiro Tsukada, and Atsutaka Maeda.
\newblock Superconductivity at 5.4 k in $\beta$-{B}i$_2${P}d.
\newblock \emph{Journal of the Physical Society of Japan}, 81\penalty0
  (11):\penalty0 113708, 2012.
\newblock \doi{10.1143/JPSJ.81.113708}.
\newblock URL \url{http://dx.doi.org/10.1143/JPSJ.81.113708}.

\bibitem[Alekseevski et~al.(1954)Alekseevski, Zhuravlev, and
  Lifanov]{Alekseevski54}
N.~Alekseevski, N.~Zhuravlev, and I.~Lifanov.
\newblock \emph{Zh. Eksp. Teor. Fiz.}, 125:\penalty0 27, 1954.

\bibitem[Shein and Ivanovskii(2013)]{Shein13}
I.~R. Shein and A.~L. Ivanovskii.
\newblock Electronic band structure and fermi surface of tetragonal
  low-temperature superconductor {B}i$_2${P}d as predicted from first
  principles.
\newblock \emph{Journal of Superconductivity and Novel Magnetism}, 26\penalty0
  (1):\penalty0 1--4, 2013.
\newblock ISSN 1557-1947.
\newblock \doi{10.1007/s10948-012-1776-x}.
\newblock URL \url{http://dx.doi.org/10.1007/s10948-012-1776-x}.

\bibitem[Sakano and et~al(2015)]{Sakano15}
M.~Sakano and et~al.
\newblock Topologically protected surface states in a centrosymmetric
  superconductor $\beta${B}i$_2${P}d.
\newblock \emph{Nature}, 6\penalty0 (8595), Octuber 2015.
\newblock \doi{10.1038/ncomms9595}.
\newblock URL
  \url{http://www.nature.com/ncomms/2015/151013/ncomms9595/pdf/ncomms9595.pdf}.

\bibitem[Coldea(2016)]{Coldea16}
A.~Coldea.
\newblock Unpublished.
\newblock 2016.

\bibitem[Ka\ifmmode \check{c}\else \v{c}\fi{}mar\ifmmode~\check{c}\else
  \v{c}\fi{}\'{\i}k et~al.(2016)Ka\ifmmode \check{c}\else
  \v{c}\fi{}mar\ifmmode~\check{c}\else \v{c}\fi{}\'{\i}k, Pribulov\'a, Samuely,
  Szab\'o, Cambel, \ifmmode~\check{S}\else \v{S}\fi{}olt\'ys, Herrera, Suderow,
  Correa-Orellana, Prabhakaran, and Samuely]{Kacmarcik16}
J.~Ka\ifmmode \check{c}\else \v{c}\fi{}mar\ifmmode~\check{c}\else
  \v{c}\fi{}\'{\i}k, Z.~Pribulov\'a, T.~Samuely, P.~Szab\'o, V.~Cambel,
  J.~\ifmmode~\check{S}\else \v{S}\fi{}olt\'ys, E.~Herrera, H.~Suderow,
  A.~Correa-Orellana, D.~Prabhakaran, and P.~Samuely.
\newblock Single-gap superconductivity in
  $\ensuremath{\beta}\text{-}\mathrm{B}{\mathrm{i}}_{2}\mathrm{Pd}$.
\newblock \emph{Phys. Rev. B}, 93:\penalty0 144502, Apr 2016.
\newblock \doi{10.1103/PhysRevB.93.144502}.
\newblock URL \url{http://link.aps.org/doi/10.1103/PhysRevB.93.144502}.

\bibitem[Herrera et~al.(2015)Herrera, Guillam\'on, Galvis, Correa, Fente,
  Luccas, Mompean, Garc\'{\i}a-Hern\'andez, Vieira, Brison, and
  Suderow]{Herrera15}
E.~Herrera, I.~Guillam\'on, J.~A. Galvis, A.~Correa, A.~Fente, R.~F. Luccas,
  F.~J. Mompean, M.~Garc\'{\i}a-Hern\'andez, S.~Vieira, J.~P. Brison, and
  H.~Suderow.
\newblock Magnetic field dependence of the density of states in the multiband
  superconductor $\beta$-{B}i$_2${P}d.
\newblock \emph{Phys. Rev. B}, 92:\penalty0 054507, Aug 2015.
\newblock \doi{10.1103/PhysRevB.92.054507}.
\newblock URL \url{http://link.aps.org/doi/10.1103/PhysRevB.92.054507}.

\bibitem[Suderow et~al.(2011)Suderow, Guillam\'on, and Vieira]{Suderow11}
H.~Suderow, I.~Guillam\'on, and S.~Vieira.
\newblock Compact very low temperature scanning tunneling microscope with
  mechanically driven horizontal linear positioning stage.
\newblock \emph{Review of Scientific Instruments}, 82\penalty0 (3):\penalty0
  033711, 2011.
\newblock URL
  \url{http://scitation.aip.org/content/aip/journal/rsi/82/3/10.1063/1.3567008}.

\bibitem[Galvis et~al.(2015)Galvis, Herrera, Guillam\'on, Azpeitia, Luccas,
  Munuera, Cuenca, Higuera, D\'iaz, Pazos, Garc\'ia-Hern\'andez, Buend\'ia,
  Vieira, and Suderow]{Galvis15}
J.A. Galvis, E.~Herrera, I.~Guillam\'on, J.~Azpeitia, R.F. Luccas, C.~Munuera,
  M.~Cuenca, J.A. Higuera, N.~D\'iaz, M.~Pazos, M.~Garc\'ia-Hern\'andez,
  A.~Buend\'ia, S.~Vieira, and H.~Suderow.
\newblock Three axis vector magnet set-up for cryogenic scanning probe
  microscopy.
\newblock \emph{Rev. Sci. Inst.}, 86:\penalty0 013706, 2015.
\newblock \doi{http://dx.doi.org/10.1063/1.4905531}.
\newblock URL
  \url{http://scitation.aip.org/content/aip/journal/rsi/86/1/10.1063/1.4905531}.

\bibitem[Rodrigo et~al.(2004)Rodrigo, Suderow, and Vieira]{Rodrigo04}
J.~G. Rodrigo, H.~Suderow, and S.~Vieira.
\newblock \emph{European Phys. Journal B}, 40:\penalty0 483, 2004.

\bibitem[Guillam\'on et~al.(2008)Guillam\'on, Suderow, Vieira, Cario, Diener,
  and Rodi\`ere]{Guillamon08}
I.~Guillam\'on, H.~Suderow, S.~Vieira, L.~Cario, P.~Diener, and P.~Rodi\`ere.
\newblock Superconducting density of states and vortex cores of
  2{H}-{N}b{S}$_2$.
\newblock \emph{Phys. Rev. Lett.}, 101:\penalty0 166407, Oct 2008.
\newblock \doi{10.1103/PhysRevLett.101.166407}.
\newblock URL \url{http://link.aps.org/doi/10.1103/PhysRevLett.101.166407}.

\bibitem[Gr\"anz et~al.(2016)Gr\"anz, Korshunov, Geshkenbein, and
  Blatter]{Granz16}
Barbara Gr\"anz, Sergey~E. Korshunov, Vadim~B. Geshkenbein, and Gianni Blatter.
\newblock Competing structures in two dimensions: Square-to-hexagonal
  transition.
\newblock \emph{Phys. Rev. B}, 94:\penalty0 054110, Aug 2016.
\newblock \doi{10.1103/PhysRevB.94.054110}.
\newblock URL \url{http://link.aps.org/doi/10.1103/PhysRevB.94.054110}.

\bibitem[Eskildsen et~al.(1998)Eskildsen, Harada, Gammel, Abrahamsen, Andersen,
  Ernst, Ramirez, Bishop, Mortensen, Naugle, Rathnayaka, and
  Canfield]{Eskildsen98}
M.~R. Eskildsen, K.~Harada, P.~L. Gammel, A.~B. Abrahamsen, N.~H. Andersen,
  G.~Ernst, A.~P. Ramirez, D.~J. Bishop, K.~Mortensen, D.~G. Naugle, K.~D.~D.
  Rathnayaka, and P.C. Canfield.
\newblock \emph{Nature}, 393:\penalty0 242--245, 1998.

\bibitem[Yethiraj et~al.(1999)Yethiraj, Christen, Paul, Miranovic, and
  Thompson]{Yethiraj99}
M.~Yethiraj, D.~K. Christen, D.~McK. Paul, P.~Miranovic, and J.~R. Thompson.
\newblock Flux lattice symmetry in ${V}_{3}\mathrm{Si}$: Nonlocal effects in a
  high- $\mathit{\ensuremath{\kappa}}$ superconductor.
\newblock \emph{Phys. Rev. Lett.}, 82:\penalty0 5112--5115, Jun 1999.
\newblock \doi{10.1103/PhysRevLett.82.5112}.
\newblock URL \url{http://link.aps.org/doi/10.1103/PhysRevLett.82.5112}.

\bibitem[Kogan et~al.(1997)Kogan, Bullock, Harmon, Miranovic-acute,
  Dobrosavljevic-acute Grujic-acute, Gammel, and Bishop]{Kogan97}
V.~G. Kogan, M.~Bullock, B.~Harmon, P.~Miranovic-acute, Lj.
  Dobrosavljevic-acute Grujic-acute, P.~L. Gammel, and D.~J. Bishop.
\newblock Vortex lattice transitions in borocarbides.
\newblock \emph{Phys. Rev. B}, 55:\penalty0 R8693--R8696, Apr 1997.

\bibitem[Fente et~al.(2016)Fente, Herrera, Guillam\'on, Suderow, Ma\~nas
  Valero, Galbiati, Coronado, and Kogan]{Fente16}
A.~Fente, E.~Herrera, I.~Guillam\'on, H.~Suderow, S.~Ma\~nas Valero,
  M.~Galbiati, E.~Coronado, and V.~G. Kogan.
\newblock Field dependence of the vortex core size probed by scanning tunneling
  microscopy.
\newblock \emph{Phys. Rev. B}, 94:\penalty0 014517, Jul 2016.
\newblock \doi{10.1103/PhysRevB.94.014517}.
\newblock URL \url{http://link.aps.org/doi/10.1103/PhysRevB.94.014517}.

\bibitem[Bending(1999)]{Bending1999}
S.~Bending.
\newblock Local magnetic pprobe of superconductors.
\newblock \emph{Advances in Physics}, 48\penalty0 (4):\penalty0 449--535, 1999.

\bibitem[Buzdin and Baladi\'e(2002)]{Buzdin02}
A.~Buzdin and I.~Baladi\'e.
\newblock Attraction between pancake vortices in the crossing lattices of
  layered superconductors.
\newblock \emph{Phys. Rev. Lett.}, 88:\penalty0 147002, Mar 2002.
\newblock \doi{10.1103/PhysRevLett.88.147002}.
\newblock URL \url{http://link.aps.org/doi/10.1103/PhysRevLett.88.147002}.

\bibitem[Koshelev(2005)]{Koshelev05}
A.~E. Koshelev.
\newblock Tilted and crossing vortex chains in layered superconductors.
\newblock \emph{Journal of Low Temperature Physics}, 139\penalty0 (1):\penalty0
  111--125, 2005.
\newblock ISSN 1573-7357.
\newblock \doi{10.1007/BF02769571}.
\newblock URL \url{http://dx.doi.org/10.1007/BF02769571}.

\bibitem[Hess et~al.(1992)Hess, Murray, and Waszczak]{Hess92}
H.F. Hess, C.A. Murray, and J.V. Waszczak.
\newblock Scanning tunneling microscopy study of distortion and instability of
  inclined flux line lattice structures in the anisotropic superconductor
  2{H}-{N}b{S}e$_2$.
\newblock \emph{Phys. Rev. Lett.}, 69:\penalty0 2138, 1992.

\bibitem[Hess et~al.(1994)Hess, Murray, and Waszczak]{Hess94}
H.~F. Hess, C.A. Murray, and J.~V. Waszczak.
\newblock Flux lattice and vortex structure in 2{H}-{N}b{S}e$_2$ in inclined
  fields.
\newblock \emph{Phys. Rev. B}, 50:\penalty0 16528, 1994.

\bibitem[Bolle et~al.(1993)Bolle, De~La~Cruz, Gammel, Waszczak, and
  Bishop]{Bolle93}
C.~A. Bolle, F.~De~La~Cruz, P.~L. Gammel, J.~V. Waszczak, and D.~J. Bishop.
\newblock Observation of tilt induced orientational order in the magnetic flux
  lattice in 2{H}-{N}b{S}e$_2$.
\newblock \emph{Phys. Rev. Lett.}, 71:\penalty0 4039--4042, Dec 1993.
\newblock \doi{10.1103/PhysRevLett.71.4039}.
\newblock URL \url{http://link.aps.org/doi/10.1103/PhysRevLett.71.4039}.

\bibitem[Gammel et~al.(1994)Gammel, Huse, Kleiman, Batlogg, Oglesby, Bucher,
  Bishop, Mason, and Mortensen]{Gammel94}
P.~L. Gammel, D.~A. Huse, R.~N. Kleiman, B.~Batlogg, C.~S. Oglesby, E.~Bucher,
  D.~J. Bishop, T.~E. Mason, and K.~Mortensen.
\newblock Small angle neutron scattering study of the magnetic flux-line
  lattice in single crystal 2 \textit{H} -${\mathrm{nbse}}_{2}$.
\newblock \emph{Phys. Rev. Lett.}, 72:\penalty0 278--281, Jan 1994.
\newblock \doi{10.1103/PhysRevLett.72.278}.
\newblock URL \url{http://link.aps.org/doi/10.1103/PhysRevLett.72.278}.

\bibitem[Fridman et~al.(2011)Fridman, Kloc, Petrovic, and Wei]{Fridman11}
I.~Fridman, C.~Kloc, C.~Petrovic, and J.Y.T. Wei.
\newblock Lateral imaging of the superconducting vortex lattice using
  doppler-modulated scanning tunneling microscopy.
\newblock \emph{Appl Phys Lett}, 99:\penalty0 192505, 2011.

\bibitem[Fridman et~al.(2013)Fridman, Kloc, Petrovic, and Wei]{Fridman13}
I.~Fridman, C.~Kloc, C.~Petrovic, and J.Y.T. Wei.
\newblock \emph{ArXiV}, page 1303.3559, 2013.

\bibitem[Kogan et~al.(1995)Kogan, Bulaevskii, Miranovi\ifmmode~\acute{c}\else
  \'{c}\fi{}, and Dobrosavljevi\ifmmode \acute{c}\else
  \'{c}\fi{}-Gruji\ifmmode~\acute{c}\else \'{c}\fi{}]{Kogan95}
V.~G. Kogan, L.~N. Bulaevskii, P.~Miranovi\ifmmode~\acute{c}\else \'{c}\fi{},
  and L.~Dobrosavljevi\ifmmode \acute{c}\else
  \'{c}\fi{}-Gruji\ifmmode~\acute{c}\else \'{c}\fi{}.
\newblock Vortex-induced strain and flux lattices in anisotropic
  superconductors.
\newblock \emph{Phys. Rev. B}, 51:\penalty0 15344--15350, Jun 1995.
\newblock \doi{10.1103/PhysRevB.51.15344}.
\newblock URL \url{http://link.aps.org/doi/10.1103/PhysRevB.51.15344}.

\bibitem[Hess et~al.(1990)Hess, Robinson, and Waszczak]{Hess90}
H.~F. Hess, R.~B. Robinson, and J.~V. Waszczak.
\newblock Vortex-core structure observed with a scanning tunneling microscope.
\newblock \emph{Phys. Rev. Lett.}, 64:\penalty0 2711, 1990.

\bibitem[et. al.((2016))]{Galvis16}
J.A.~Galvis. et. al.
\newblock \emph{In preparation}, (2016).

\bibitem[M\"uhlbauer et~al.(2009)M\"uhlbauer, Pfleiderer, B\"oni, Laver,
  Forgan, Fort, Keiderling, and Behr]{Muhlbauer09}
S.~M\"uhlbauer, C.~Pfleiderer, P.~B\"oni, M.~Laver, E.~M. Forgan, D.~Fort,
  U.~Keiderling, and G.~Behr.
\newblock Morphology of the superconducting vortex lattice in ultrapure
  niobium.
\newblock \emph{Phys. Rev. Lett.}, 102:\penalty0 136408, Apr 2009.
\newblock \doi{10.1103/PhysRevLett.102.136408}.
\newblock URL \url{http://link.aps.org/doi/10.1103/PhysRevLett.102.136408}.

\end{thebibliography}

\end{document}


\title{Supplemental material for "{Subsurface bending and reorientation of the tilted vortex lattice in the bulk    due to  Coulomb-like   repulsion at the surface}"} 

\author{E. Herrera}
\affiliation{Laboratorio de Bajas Temperaturas y Altos Campos Magn\'eticos, Unidad Asociada UAM/CSIC, Departamento de F\'isica de la Materia Condensada, Instituto de Ciencia de Materiales Nico\'las Cabrera, Instituto de F\'isica de la Materia Condensada, Universidad Aut\'onoma de Madrid, E-28049 Madrid, Spain}

\author{I. Guillam\'on}
\affiliation{Laboratorio de Bajas Temperaturas y Altos Campos Magn\'eticos, Unidad Asociada UAM/CSIC, Departamento de F\'isica de la Materia Condensada, Instituto de Ciencia de Materiales Nico\'las Cabrera, Instituto de F\'isica de la Materia Condensada, Universidad Aut\'onoma de Madrid, E-28049 Madrid, Spain}

\author{J.A. Galvis}
\affiliation{Laboratorio de Bajas Temperaturas y Altos Campos Magn\'eticos, Unidad Asociada UAM/CSIC, Departamento de F\'isica de la Materia Condensada, Instituto de Ciencia de Materiales Nico\'las Cabrera, Instituto de F\'isica de la Materia Condensada, Universidad Aut\'onoma de Madrid, E-28049 Madrid, Spain}
\affiliation{Departamento de ciencias naturales, Facultad de ingenieria, Universidad Central, Bogot\'a, Colombia.}
\affiliation{National High Magnetic Field Laboratory, Florida State University, Tallahassee, FL 32310, USA.}

\author{A. Correa}
\affiliation{Laboratorio de Bajas Temperaturas y Altos Campos Magn\'eticos, Unidad Asociada UAM/CSIC, Departamento de F\'isica de la Materia Condensada, Instituto de Ciencia de Materiales Nico\'las Cabrera, Instituto de F\'isica de la Materia Condensada, Universidad Aut\'onoma de Madrid, E-28049 Madrid, Spain}
\affiliation{Instituto de Ciencia de Materiales de Madrid, Consejo Superior de Investigaciones Cient\'ificas, CSIC, E-28049 Madrid, Spain}

\author{A. Fente}
\affiliation{Laboratorio de Bajas Temperaturas y Altos Campos Magn\'eticos, Unidad Asociada UAM/CSIC, Departamento de F\'isica de la Materia Condensada, Instituto de Ciencia de Materiales Nico\'las Cabrera, Instituto de F\'isica de la Materia Condensada, Universidad Aut\'onoma de Madrid, E-28049 Madrid, Spain}

\author{S. Vieira}
\affiliation{Laboratorio de Bajas Temperaturas y Altos Campos Magn\'eticos, Unidad Asociada UAM/CSIC, Departamento de F\'isica de la Materia Condensada, Instituto de Ciencia de Materiales Nico\'las Cabrera, Instituto de F\'isica de la Materia Condensada, Universidad Aut\'onoma de Madrid, E-28049 Madrid, Spain}

\author{H. Suderow}
\affiliation{Laboratorio de Bajas Temperaturas y Altos Campos Magn\'eticos, Unidad Asociada UAM/CSIC, Departamento de F\'isica de la Materia Condensada, Instituto de Ciencia de Materiales Nico\'las Cabrera, Instituto de F\'isica de la Materia Condensada, Universidad Aut\'onoma de Madrid, E-28049 Madrid, Spain}

\author{A. Yu. Martynovich}
\affiliation{Ames Laboratory and Department of Physics \& Astronomy, 
Iowa State University, Ames, Iowa 50011, USA}

\author{V. G. Kogan}
\affiliation{Ames Laboratory and Department of Physics \& Astronomy, 
Iowa State University, Ames, Iowa 50011, USA}

\begin{abstract}
We provide additional data for low tilt angles and explicit expressions for the intervortex  distances. The   nearest neighbor intervortex distances at the surface  are   equal or larger than the bulk intervortex distance.
\end{abstract}

\maketitle

\section{Vortex lattices at the surface}
\vspace{-0.2cm}
Within the  {\it isotropic} model, the VL in the vortex frame of an infinite sample is hexagonal and {\it degenerate}: the angle $\alpha$ shown in the upper panel of Fig.\,2(c) of the main text can be taken as the degeneracy parameter. For a given $\alpha$, the VL unit cell vectors (in units of $a$) are given by the Eq.\,(1). Positions of the nearest neighbors are $\pm \bm u_1$, $\pm \bm u_2$, and $\pm(\bm u_1- \bm u_2)$, all of them at the same distance $d=1$ from the vortex at the origin.

It is argued in the main text that considering the VL structure, one can disregard the vortex bending. Then,  the bulk nearest neighbors will be situated at the crossection of a circular cylinder of radius $a$ with the  surface plane, i.e. at the ellipse with semi-axes $a/\cos\theta$ and $a$ (right panel of Fig.\,2c). Taking the $x$ axis of the surface frame in the tilt plane, one obtains new unit cell vectors at the surface  (in units of $a$) given by Eq.\,(2).

The positions corresponding to the bulk nearest neighbors are $ \pm\bm v_1$, $ \pm\bm v_2$, and  $ \pm(\bm v_1-\bm v_2)$. These positions at the surface are not necessarily the nearest. Their distances from the vortex at the origin are 
  $d_1=|\bm v_1|$, $d_2=|\bm v_2|$, and $d_3=|\bm v_1-\bm v_2|$.  For these distances one has:
        \begin{eqnarray}
 d_1^2&=& \frac{ \cos^2\alpha +\cos^2\theta\sin^2\alpha}{\cos^2\theta} \,,\label{d1}\\
  d_2^2&=& \frac{ \cos^2(\alpha+\pi/3) +\cos^2\theta\sin^2(\alpha+\pi/3)}{\cos^2\theta}\,,\qquad\label{d2}\\
 d_3^2&=& \frac{ [\cos (\alpha+\pi/3) -\cos\alpha]^2}{ \cos^2\theta}+
  \left[\sin \left(\alpha+\frac{\pi}{3}\right)-\sin\alpha\right]^2  .\nonumber\\
 \label{d3}
 \end{eqnarray}

Consider now the case $\alpha=\pi/3$ (which is equivalent to $\alpha=0$, the structure $A^\prime$), see Fig.\,2c. With increasing tilt $\theta$, the position  $  (\bm v_1+\bm v_2)$, which at small $\theta$ does not belong to the nearest neighbors,  approaches the origin, whereas the distance to $  (\bm v_1-\bm v_2)$ increases. The two distances become equal when $  (\bm v_1+\bm v_2)^2= (\bm v_1-\bm v_2)^2$, in other words when $  \bm v_1\cdot\bm v_2=0$, i.e. $  \bm v_1\perp\bm v_2$. Since in this case by symmetry $  |\bm v_1|=|\bm v_2|$, the surface VL will appear as having the square unit cell. A simple algebra yields for the tilt $\theta^*$ at which this happens: $\cos^2\theta^*=-\cot(2\pi/3)/\sqrt{3}$ which corresponds to  $\theta^*=70.5^\circ$. We observe the square VL at angles that are close to that, namely $\theta=80^\circ$ and $\alpha=25^\circ$.
\section{Low polar angles}

\begin{figure*}[h]
\includegraphics[width=0.9\textwidth]{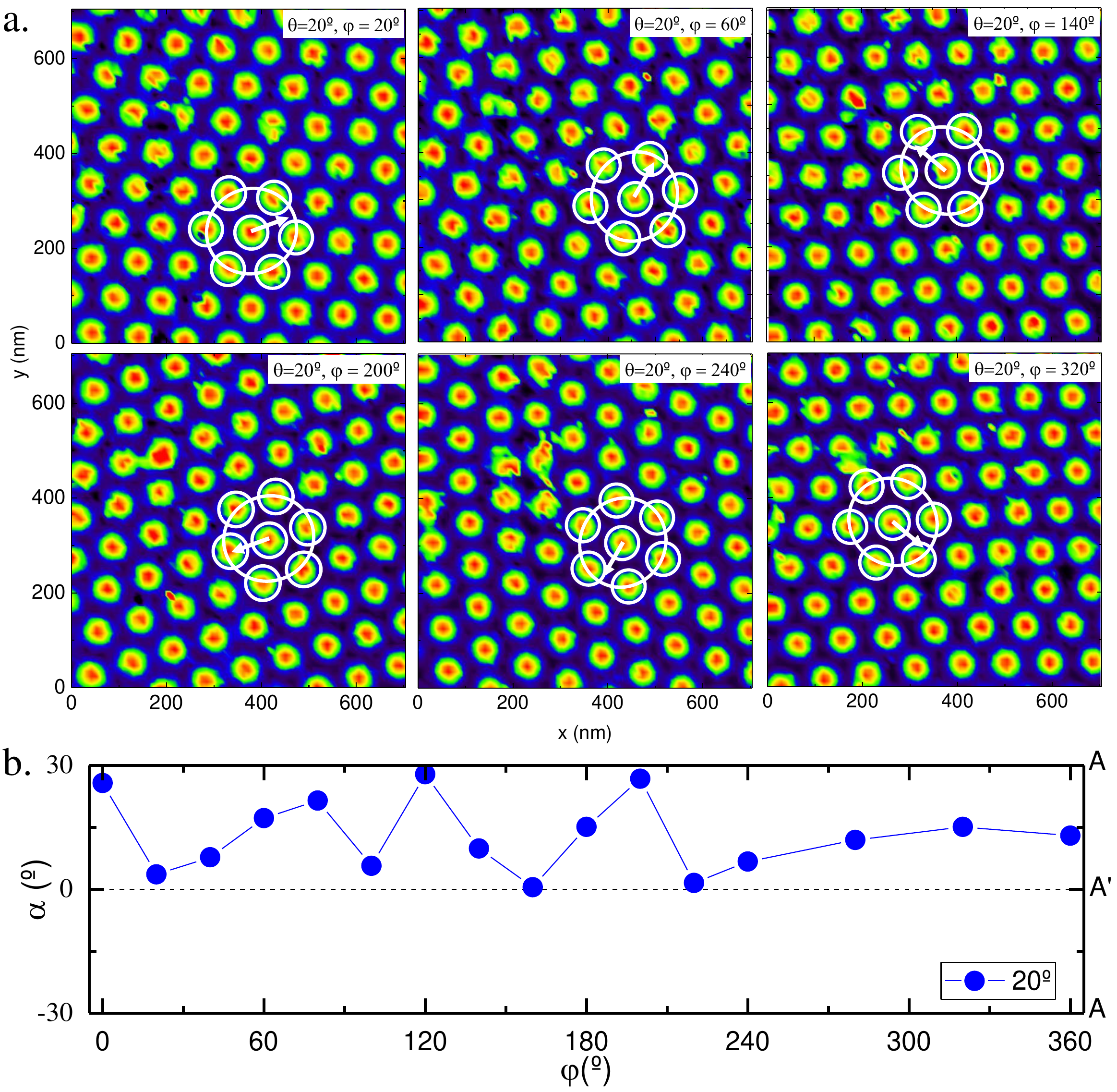}
\caption{(Color online). As in Fig. 4 of the main text, VLs for a fixed tilt $\theta=20^{\circ}$ but different azimuthal angles $\varphi$ of the applied field. White ellipses are drawn according to Fig. 2c and enclose the nearest neighbor positions (circles) of the surface vortex lattice. White arrows represent the in-plane magnetic field projection.}
\label{Fig1S}
\end{figure*}

From Fig.\,3 of the publication, one can see  that the structure $A$ ($\alpha=-\pi/6$) is unstable since the energy has a maximum at this configuration. Moreover, this feature clearly does not depend on the number of neighbors included in the summation. Hence, we expect that for relatively small tilts, minimum energy shown in Fig.\,3 corresponds to small $\alpha$ (along with $\alpha_S$) i.e., the VL structure is close to $A^\prime$. For $\theta=0.5\approx 28.6^\circ$, the distances $d_i$ are $1.14,\,\,1.04,\,\,1.04$, all three are larger than 1, as expected because of the Coulomb repulsion. Note that for the lattice $A$, there will be one of the $d_i$ equal to $1$. Thus, even the smallest intervortex distance at the surface exeeds the bulk intervortex distance $d_0$. 

We examine the situation for a  polar angle $\theta=20^{\circ}$. The result of the experiment is shown in Fig.\ref{Fig1S}a. Here, the vortex lattice remains practically undistorted. In Fig.\ref{Fig1S}b we show $\alpha$ for $\theta=20^{\circ}$ and different azimuthal angles $\varphi$. As we see in Fig.\ref{Fig1S}, the angle $\alpha$ corresponds to a structure different from $A$ and $A'$. We believe that this difference might be due to pinning. This question should be further studied.